\documentclass[pra,aps,twocolumn,eqsecnum,showpacs]{revtex4}
\usepackage{mathtext}
\usepackage[T2A]{fontenc}
\usepackage[cp1251]{inputenc}
\usepackage{epsf}
\usepackage{psfrag}
\usepackage{graphicx}
\usepackage{amssymb,amsmath}

\begin{document}
\title{Angular distribution of single photon superradiance in a dilute and cold atomic ensemble}
\author{A. S. Kuraptsev${}^{1}$ and I. M. Sokolov${}^{1,2}$ \\
{\small $^{1}$Department of Theoretical Physics, Peter the Great St. Petersburg Polytechnic University, 195251, St.-Petersburg, Russia}\\
{\small $^{2}$Institute for Analytical Instrumentation, Russian Academy of Sciences, 198095, St.-Petersburg, Russia }\\
M. D. Havey\\
{\small Department of Physics, Old Dominion University, Norfolk, VA 23529}}

\begin{abstract}
On the basis of a quantum microscopic approach we study the dynamics of the afterglow of a dilute Gaussian atomic ensemble excited by pulsed radiation. Taking into account the
vector nature of the electromagnetic field we analyze in detail the angular and polarization distribution of single-photon superradiance of a such ensemble. The dependence of
the angular distribution of superradiance on the length of the pulse and its carrier frequency as well as on the size and the shape of the atomic clouds is studied. We show that
there is substantial dependence of the superradiant emission on the polarization and the direction of fluorescence. We observe essential  peculiarities of superradiance in the region of
the forward diffraction zone and in the area of the coherent backscattering cone. We demonstrate that there are directions for which the rate of fluorescence is several times
more than the decay rate of the timed-Dicke state. We show also that single-photon superradiance can be excited by incoherent excitation when atomic polarization in the ensemble
is absent. Besides a quantum microscopic approach, we analyze single-photon superradiance on the basis of the theory of incoherent multiple scattering in optically thick media
(random walk theory).  In the case of very short resonant and long nonresonant pulses we derive simple analytical expressions for the decay rate of single-photon superradiance
for incoherent fluorescence  in an arbitrary direction.
\end{abstract}


\maketitle
\section{Introduction}

Since the original work by Dicke \cite{1}, the problem of superradiance, and it counterpart subradiance, have attracted great interest. By the end of the 1980's many aspects of
the physics of the superradiance problem had been studied in detail (see \cite{2} and references therein). However, in the past decade, theoretical and experimental advances
have led to a rejuvenation of this field.  In particular, theoretical predictions \cite{3,4} and experimental studies have been made of single photon superradiance and the
collective Lamb shift in the X-ray regime \cite{5}, in cold atoms \cite{6,7,7a}, and in quantum dots \cite{7b}.

Contrary to the "traditional"\, superradiance predicted by Dicke for polyatomic ensembles with sizes much smaller than the radiation wavelength, the single-photon superradiance
is observed for large-sized systems. It differs also from well studied superfluorescence of long and extended ensembles with a large number of initially excited atoms.
Single-photon superradiance is a linear-optics effect and it can take place for dilute atomic systems under excitation by a weak pulse of radiation. By now the main features of
these effects have been studied both theoretically and experimentally \cite{5,6,7,7a,7b} (and references therein).

Due to the effect of coherent forward scattering, the main part of the radiation pulse absorbed by the extended atomic ensemble is scattered into directions close to that of the
exciting light pulse \cite{3,8}. In this connection the main attention was given to the properties of superradiance emitted in the forward direction. Particularly it was shown
that one way to obtain strong coherent emission from the ensembles  is to prepare a timed-Dicke state \cite{9}. However, as was shown in \cite{10} and studied in a sophisticated
experiment \cite{7} superradiance can be observed in directions outside the bounds of the main diffraction cone. Moreover, the fluorescence decay rate in these directions can
exceed the decay rate of the timed-Dicke state responsible for forward radiation. In several works \cite{7,10} the time dependence of fluorescence in certain, fixed directions
was studied. The main goal of the present paper is to study theoretically the angular distribution of superradiance in more detail. Among other things, we will show that
superradiance is characterized by a strong and essentially nonmonotonic angular dependence.

In analysis of the angular dependence of single-photon superradiance, the polarization properties of the fluorescence may play an important role. In experiments \cite{6,7} the
total intensity of scattered light was measured.  The polarization dependence was not analyzed in detail. For radiation coherently scattered into the forward direction  the
polarization of superradiance should coincide with that of incident light. In \cite{6} it was experimentally confirmed  in the case of the linear polarization of the excitation
pulse. So the polarization need not be studied. In the case of sideways scattering it is not the case. Decay of fluorescence in different polarization channels has in the general
case different rates. For this reason, in studying  single-photon superradiance, we will make additional analysis of its polarization properties.

Another goal of the present work is to study the influence of the type of excitation on single-photon superradiance. For scattering in the main diffraction lobe this influence
has been studied in great length. Particularly it has been discussed how the nature of superradiance changes if the initial state differs from a timed-Dicke state. In this work
we will consider the influence of the pulse duration on the angular distribution of superradiance. Broadly speaking, single photon excitation is not necessary for superradiance.
In the paper \cite{11} it was shown that the initial spatially extended atomic coherence is important. In real optical experiments excitation of the atomic ensemble is performed
by means of a light pulse. In experiment \cite{6} it was a short pulse with a length less than the lifetime of the excited states of the free atom. In \cite{7} the pulse
duration exceeded essentially the lifetime. For a large detuning of the carrier frequency of the pulse the ensemble is optically thin and all atoms were excited with the same
probability as in a timed-Dicke state. At the same time the scattering at a large angle is incoherent and this raises the question of whether coherent excitation is necessary
for observation of single-photon superradiance beyond the main diffraction lobe. In this work we consider not only the dependence of the angular distribution of the
superradiance on the length of the exciting pulse but also the possibility to observe superradiance in the case of noncoherent excitation. We will show that it is indeed
possible.

Finally, we will study the dependence of the angular distribution of the decay rate on the shape of the atomic ensemble. We will consider how this distribution is modified by
the change of the aspect ratio of an elliptical sample with a Gaussian density distribution.

\section{Basic assumptions and approach}

In our calculations of time-dependent fluorescence we  will follow the theoretical approach developed previously in \cite{12}. In the framework of this approach we solve the
nonstationary Schrodinger equation for the wave function $\psi$ of the joint system consisting of all atoms and a weak electromagnetic field. A vacuum reservoir is also included
in our considerations.

We consider a disordered atomic cloud of N  two-level atoms. All atoms have a ground state $|g\rangle$ with the total angular momentum $J_g = 0$, an excited state $|e\rangle$
with $J_e = 1$, a transition frequency $\omega_a$, and a natural lifetime of the excited state $\tau_0 = 1/\gamma$. Taking into account the experimentally relevant situation of
a cold atomic cloud we assume atoms to be motionless and located at random positions $\mathbf{r}_i,\,  (i = 1, . . . ,N)$. Possible atomic displacement caused by residual atomic
motion is taken into account by averaging of calculated quantities over random spatial distribution of the atoms.

We seek the wave function $\psi$ as an expansion in a set of eigenfunctions of the Hamiltonian $H_0$ of the noninteracting atoms and field. The key simplification of the
approach employed is in the  restriction of the total number of states taken into account. Assuming that the exciting radiation is weak, which is typical in experiments
\cite{6,7}, we take into account only states with no more than one photon in the field. As it was shown in  \cite{13}-\cite{16}, this approximation allows us to describe
collective effects under scattering of weak radiation, including pulsed radiation.

Knowledge of the wave function gives us information about the properties of the atomic ensemble as well as the properties of the secondary radiation. In particular, the
intensity $I_\alpha (\mathbf{\Omega},t )$ of the light  polarization component $\alpha$ that the atoms scatter in a unit solid angle around an arbitrary direction given by
radius-vector $\mathbf{r}$ ($\mathbf{\Omega}={\theta,\varphi}$) can be determined as follows
\begin{equation} I_\alpha (\mathbf{\Omega},t )=\frac{c}{4\pi}  \left\langle \psi
\right\vert E^{(-)}_\alpha(\mathbf{r}) E^{(+)}_\alpha(\mathbf{r}) \left\vert \psi \right\rangle r^2. \label{2}
\end{equation}
Here $E^{(\pm)}_\alpha(\mathbf{r})$ are the positive and negative frequency parts of the electric field operator.

In the case of pulsed excitation, the mean value in this expression depends on time. The corresponding dependence can be found as the inverse Fourier transform (for more details
see \cite{17})
\begin{eqnarray}
\left\langle \psi \right\vert E^{(-)}_\alpha(\mathbf{r}) E^{(+)}_\alpha(\mathbf{r}) \left\vert \psi \right\rangle = \left\vert\int\limits_{-\infty }^{\infty
}\dfrac{\hbar\exp(-i\omega t)d\omega }{2\pi } \right. \notag  \\ \left. \sum\limits_{e,e^{\prime }}\widetilde{\Sigma }_{\alpha e}(\omega )R_{ee^{\prime }}(\omega )
\Lambda_{e^{\prime }}(\omega ) \right\vert ^{2}.  \label{3}
\end{eqnarray}
Here the vector $ \Lambda_{e}(\omega )$ describes excitation of different states of different atoms by external pulsed radiation
\begin{equation} \Lambda_{e}(\omega)=-\frac{\mathbf{d}_{e;g}\mathbf{E}(\omega)}{\hbar }=
-\frac{\mathbf{u}\mathbf{d}_{e;g}}{\hbar }E_0(\omega)\exp(i\mathbf{kr_e}), \label{4}
\end{equation}
In this equation $\mathbf{d}_{e;g}$ is the dipole matrix element for the transition from the ground $g$ to the excited $e$ state of the atom, $E_{0}(\omega)$ is a Fourier
amplitude of the probe radiation, which we assume to be a plane wave, $\mathbf{k}$ and $\mathbf{u}$ are its wave vector and unit polarization vector, $\mathbf{r}_{e}$ is the
radius-vector of the atom $e$.

The matrix $R_{ee^{\prime }}(\omega )$ is the resolvent of the considered system projected on the one-fold atomic excited states
\begin{eqnarray}
R_{ee^{\prime }}(\omega )=\left[ (\omega -\omega _{e})\delta _{ee^{\prime }}-\Sigma _{ee^{\prime }}(\omega)\right] ^{-1}.  \label{5}
\end{eqnarray}
In this work we determine it numerically on the basis of the known expression for the matrix $\Sigma _{ee^{\prime }}(\omega)$. Matrix elements $\Sigma _{ee^{\prime }}(\omega )$
for $e$ and $e^\prime$ corresponding to different atoms describe excitation exchange between these atoms
\begin{eqnarray}
\label{6} &&\Sigma _{ee^{\prime }}(\omega )=\sum\limits_{\mu ,\nu} \frac{\mathbf{d}_{e_{a};g_{a}}^{\mu }\mathbf{d}_{g_{b};e_{b}}^{\nu }}{\hbar r^{3}}\times \\&& \left[ \delta
_{\mu \nu }\left( 1-i\frac{\omega _{a}r}{c}-\left( \frac{\omega _{a}r}{c}\right) ^{2}\right) \exp \left( i\frac{\omega _{a}r}{c}\right) \right. - \notag  \\&& \left.
-\dfrac{\mathbf{r}_{\mu }\mathbf{r}_{\nu }}{r^{2}}\left( 3-3i\frac{\omega _{a}r}{c}-\left( \frac{\omega _{a}r}{c}\right) ^{2}\right) \exp \left( i\frac{\omega _{a}r}{c}\right)
\right]. \notag
\end{eqnarray}
This expression is written assuming that in states $\psi _{e^{\prime }}$  and $\psi _e$ atoms $b$  and $a$ are excited correspondingly, we used also the pole approximation
($\Sigma _{ee^{\prime }}(\omega )=\Sigma _{ee^{\prime }}(\omega_a )$, see \cite{18}). In (\ref{6}) $\mathbf{r}_\mu$ is projections of the vector
$\mathbf{r}=\mathbf{r}_{a}-\mathbf{r}_{b}$ on the axes of the chosen reference frame and  $r=|\mathbf{r}|$ is the spacing between atoms  $a$ и $b$.

If $e$ and $e^\prime$ correspond to excited states of one atom then $\Sigma _{ee^{\prime }}(\omega )$ differs from zero only for $e=e^\prime$ (i.e. $m=m^\prime$, where $m$ is
magnetic quantum number of the atomic excited state). In this case
\begin{equation} \Sigma _{ee}(\omega )=-i\gamma/2.  \label{7}
\end{equation}

The matrix $\widetilde{\Sigma }_{\alpha e}(\omega )$  in (\ref{3}) describes light propagation from an atom excited in the state $e$ to the photodetector. In the rotating wave
approximation it is (see \cite{12})
\begin{eqnarray}
\label{8} \widetilde{\Sigma }_{\alpha e}(\omega)=-\frac{\mathbf{u'}_{\alpha }^{\ast}\mathbf{d}_{g;e}}{\hbar r } \left( \frac{\omega }{c}\right) ^{2}\exp \left( i\frac{\omega
\left\vert\mathbf{r-r}_{e}\right\vert }{c}\right)  \\ \notag \approx
 -\frac{\mathbf{u'}_{\alpha }^{\ast}\mathbf{d}_{g;e}}{\hbar r } \left( \frac{\omega
}{c}\right) ^{2}\exp \left( i\frac{\omega r }{c}-i\frac{\mathbf{k'r}_e}{c}\right) .
\end{eqnarray}
Here $\mathbf{u'}_{\alpha }^{\ast}$ is a unit polarization vector of the scattered wave and $\mathbf{k'}$ is its wave vector.

Substituting  (\ref{4}) and  (\ref{8}) into  (\ref{3}), after some simplifications we have
\begin{eqnarray}
\label{9}
&&I_\alpha(\mathbf{\Omega},t )=\frac{c}{4\pi\hbar^2} \left\vert\int\limits_{-\infty }^{\infty }E_0(\omega)k^{2}\dfrac{\exp(-i\omega t)d\omega }{2\pi } \right. \\
&& \sum\limits_{e,e^{\prime }}\left.\left(\mathbf{u}^{\prime \ast }\mathbf{d}_{g;e}\right) R_{ee^{\prime }}(\omega ) \left(\mathbf{ud}_{e^{\prime };g}\right) \exp \left(
i(\mathbf{kr}_{e^{\prime }}-\mathbf{k}^{\prime}\mathbf{r}_{e})\right)  \right\vert ^{2}. \notag
\end{eqnarray}
The total intensity $I(\mathbf{\Omega},t) $ can be obtained as a sum of (\ref{9}) over two orthogonal polarizations $\alpha$.

Note that approaches similar to those described in this paper are used to analyze the atomic decay or dynamics of  fluorescence in several works \cite{18}-\cite{17j}. In the
main part of the mentioned references, the  scalar approximation was used. It is known that for the dilute clouds we are interested in here this approach is quite appropriate
for description of a whole series of properties \cite{18a}-\cite{18c} (if one takes into account that it underestimates optical thickness, see below). However, in the present
work we are going to study the polarization dependence of fluorescence and we have to avoid this simplification.

In the next section, we will use relation (\ref{9})  to analyze temporal, polarization and angular properties of the  scattered light.

\section{Results and discussion}
In the framework of our approach we can analyze atomic ensembles with arbitrary shape and spatial distribution of atoms. In the present work we will consider axially symmetric
Gaussian clouds having an average density distribution given by
\begin{equation}
n(\mathbf{r})=n_0 \exp\left(-\frac{z^2}{2L^2}-\frac{x^2+y^2}{2R^2}\right).
 \label{1}
\end{equation}
The incident light is a plane wave propagating in the z direction except for the case when we consider incoherent excitation. In the latter case we will consider quasi isotropic
irradiation from all directions. For illustrative purposes, we will restrict our consideration to temporally rectangular pulses having a central frequency $\omega_L$. The length
of the pulse is $\tau_L$. We will assume that the zero-time reference $t = 0$ corresponds to the end of the exciting pulse. In all calculations the incident light is left-handed
circularly polarized.

In the following we will use Eq. (\ref{9}) averaged over the ensemble of possible atomic configurations to study the average intensity of the time dependent fluorescence
$\langle I_\alpha(\mathbf{\Omega},t )\rangle $. Collective effects not only accelerate the fluorescence in some directions  but also modify the functional form of $\langle
I_\alpha(\mathbf{\Omega},t )\rangle $ making it nonexponential. To analyze the peculiarities of the time dependence we introduce a current decay rate as follows
\begin{equation} \Gamma_\alpha(\mathbf{\Omega},t)=\frac{\partial ln\langle I_\alpha(\mathbf{\Omega},t )\rangle }{\partial t}.  \label{10}
\end{equation}
This rate is a function of time $t$. For the total fluorescence without polarization analysis we will use a similar relation but with summed intensity $\langle
I(\mathbf{\Omega},t )\rangle  =\sum_\alpha\langle I_\alpha(\mathbf{\Omega},t )\rangle $.

\subsection{Angular distribution of scattered light}

In Fig. 1 we show the angular distribution of light scattered in different polarization channels.  The calculation is performed for a spherically symmetric atomic ensemble. The
radius of the Gaussian distribution is $R=L=25\lambdabar$. Hereafter in this paper we use $\lambdabar$  as a unit of length, where $\lambdabar$ = $\lambda$/$2\pi$. The peak
density is $n_0=0.005$. Fig. 1a corresponds to a time equal to $t=\tau_0$ after the exciting pulse is switched off. It demonstrates very different behaviors for different
polarizations of the scattered light. In the helicity preserving channel ($H\|H$) we see a typical diffraction picture. There is large main diffraction peak. Two higher order
peaks are also well distinguished, although because of a smooth decrease in the concentration at the edges of the Gaussian cloud, these peaks are not as distinct as for
diffraction from objects with sharp boundaries. The scattering into the back half-sphere is suppressed in this polarization channel. For the helicity-nonpreserving ($H\bot H$)
polarization channel the main part of the radiation is scattered into the backward direction. The intensity is mainly determined by single scattering from the boundary region.
For the $H\|H$ channel single scattering in the exact backward direction is absent because of selection rules for atomic electric dipole transitions.
\begin{figure}[th]
\begin{center}
\includegraphics[width=19.5pc]{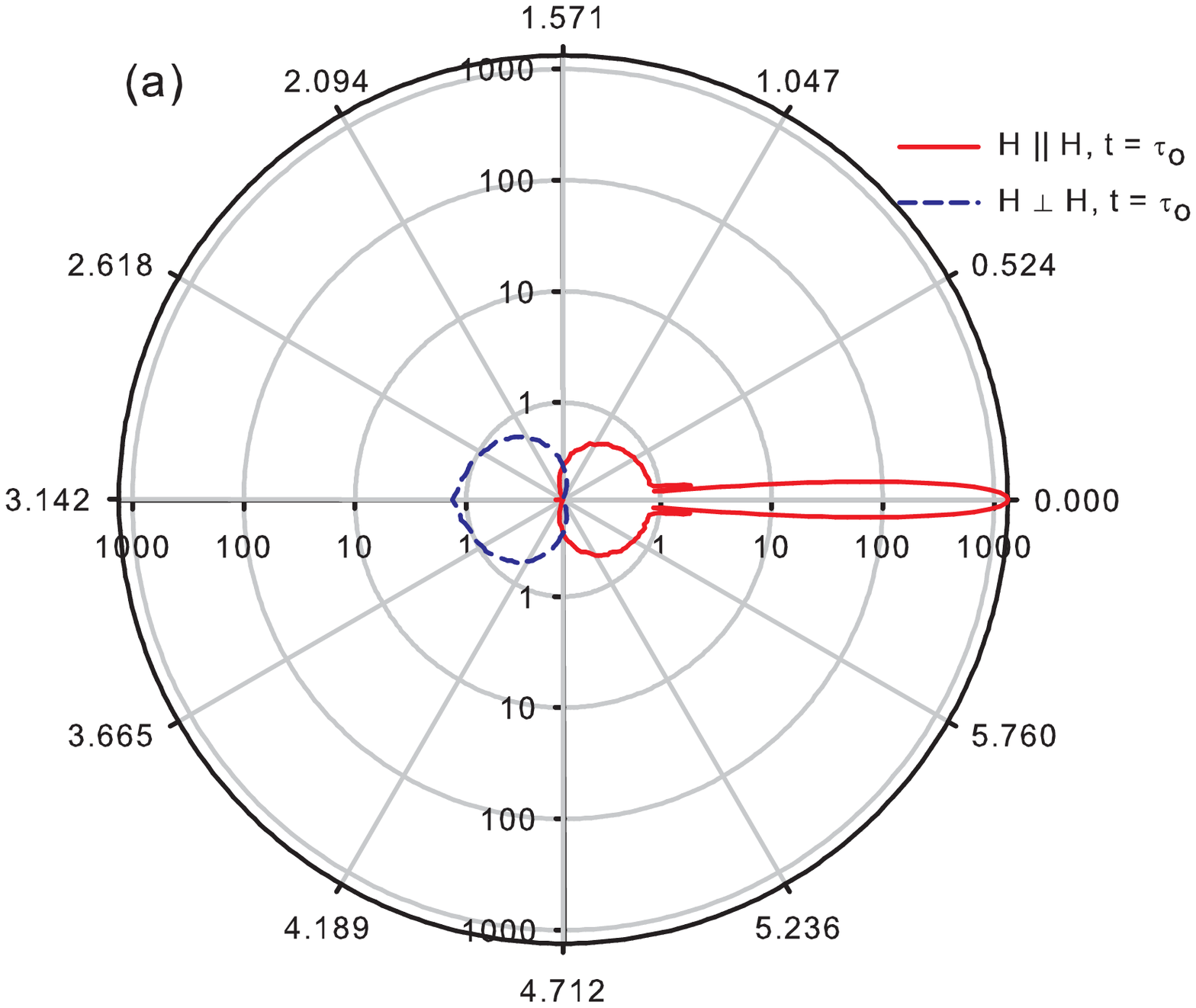}
\includegraphics[width=19.5pc]{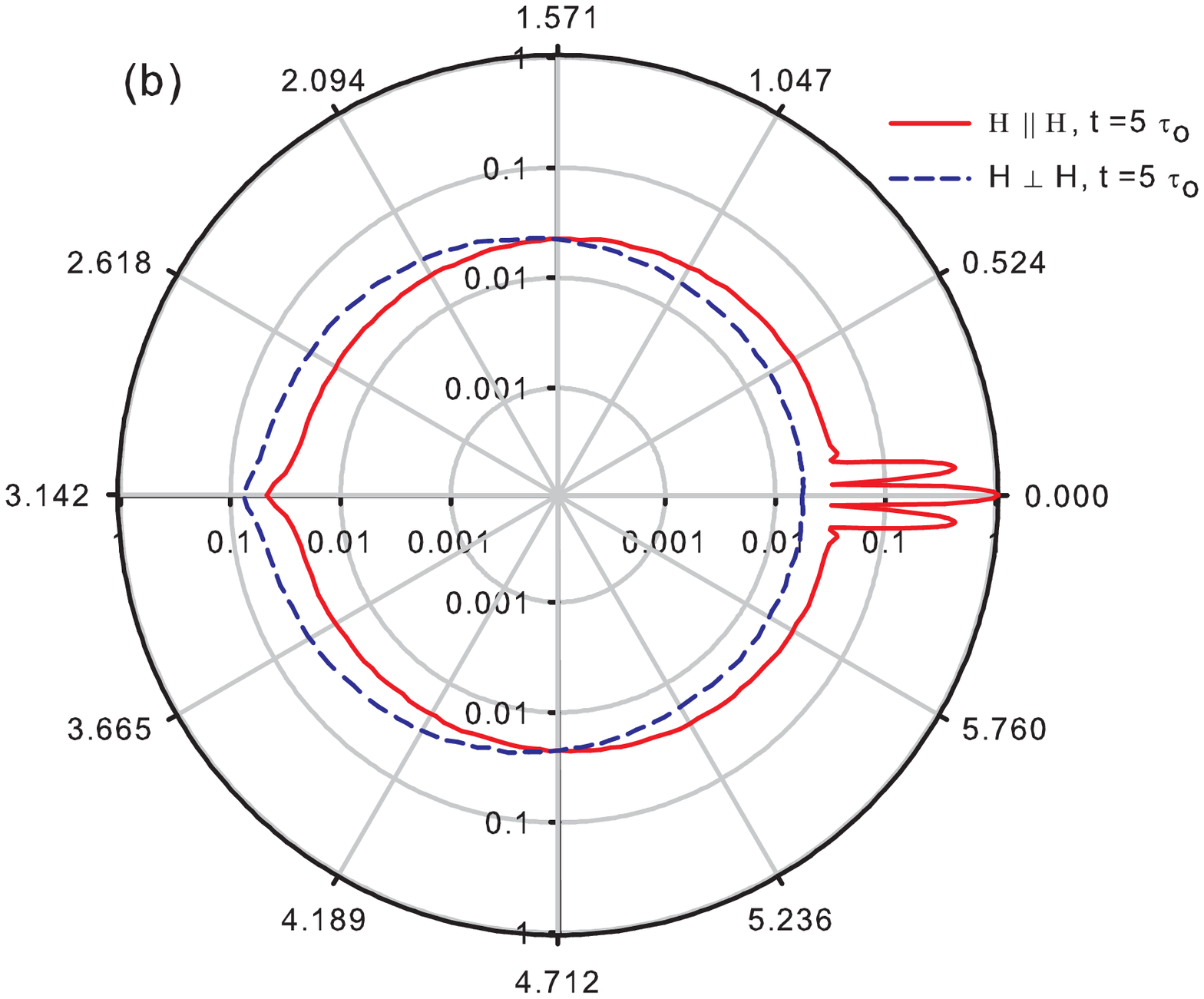}
\end{center}
\caption{Angular distribution of light scattered in different polarization channels for different times after the excitation pulse is switched off. (a) $t=\tau_0$; (b)
$t=5\tau_0$. Spherically symmetric Gaussian cloud $L=R=25\lambdabar$, $n_0\lambdabar^{3}=0.005.$ Pulse length is $\tau_L=0.1\tau_0$
 } \label{f1}
\end{figure}

The angular distribution of fluorescence changes with time. In Fig. 1b this effect is shown for $t=5\tau_0$. For this time, the main contribution to the fluorescence is
determined by multiply scattered light. The difference between polarization channels becomes less evident. Further, the angular distribution becomes more spherically symmetric
in each channel.  However even at this time interval we see some traces of a diffraction picture. In both channels we see also the cone-shaped feature associated with coherent
backscattering (see for example reviews \cite{19,20} and references therein). The enhancement factor for the helicity preserving channel is close to two which is typical for a
0-1 transition. For another channel it is much less because of the single scattering contribution to the background \cite{19,20}.

We calculated angular dependencies like those shown in Fig.1 for different instants of time and thus determined the current decay rate (\ref{10}) for fluorescence in any
direction and for any polarization channels. Consider at first, however, the decay rate for the total intensity as it was made in experiments \cite{6,7}.

In Fig. 2 we show the angular dependence of $\Gamma(\mathbf{\Omega},t)$ averaged over some time intervals $\Delta t$. For a clearer demonstration we displaced the graphs along
the abscissa.
\begin{figure}[th]
\begin{center}
\includegraphics[width=18pc]{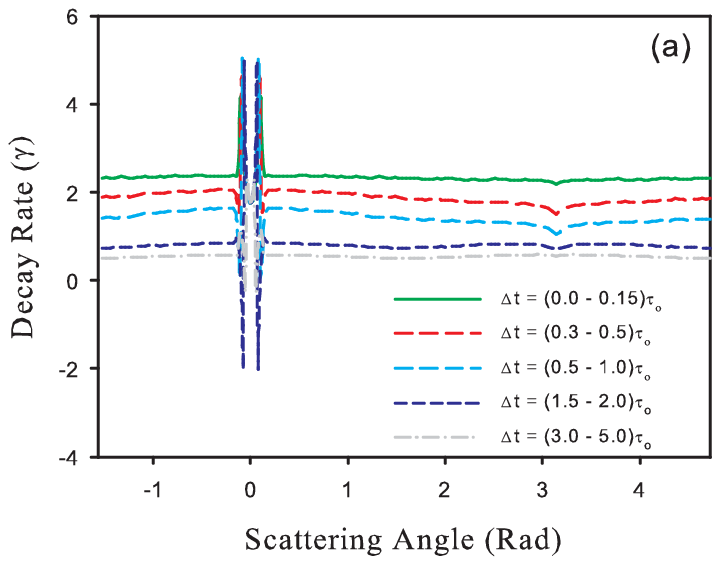}
\includegraphics[width=18pc]{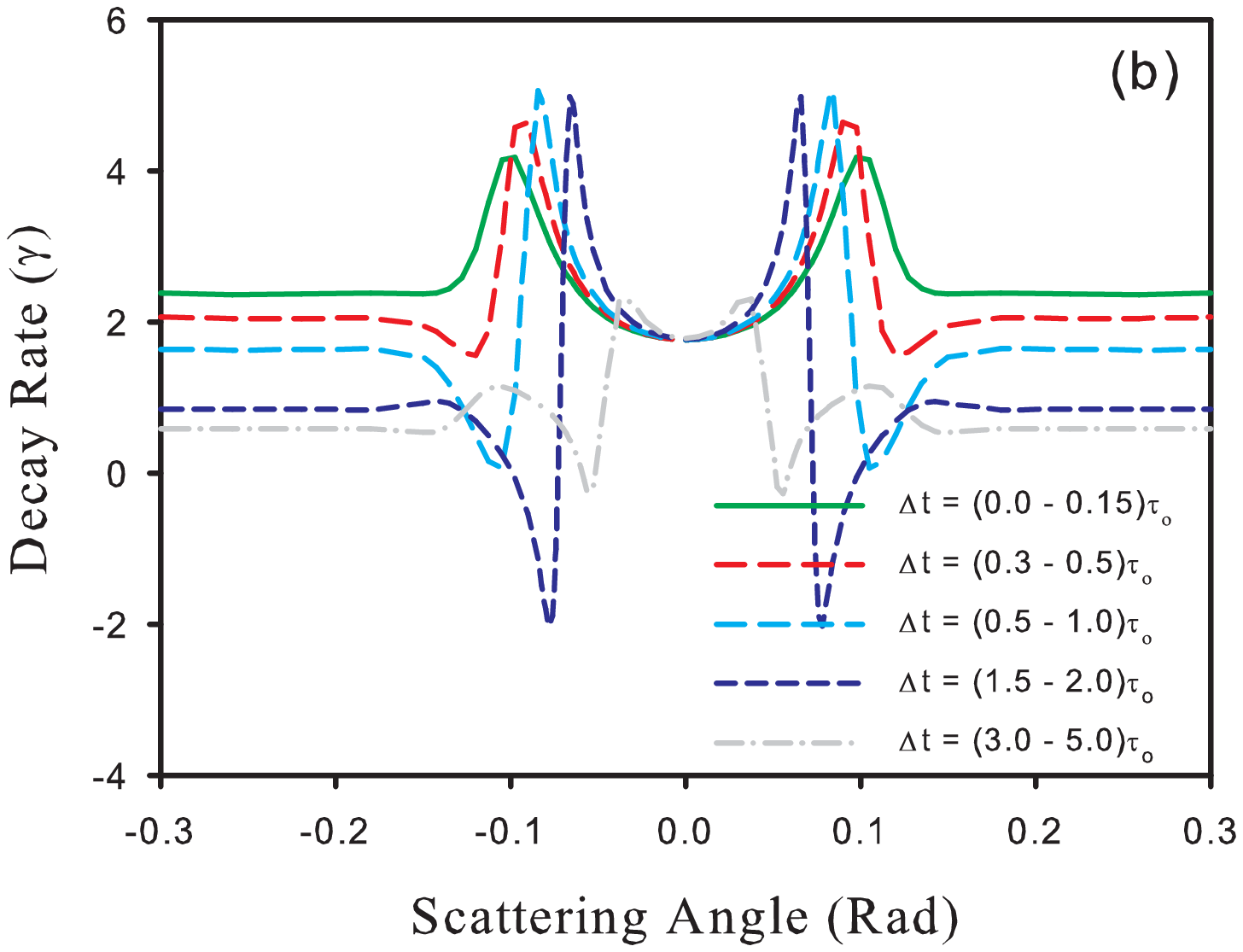}
\end{center}
\caption{Angular distribution of averaged decay rate $\Gamma(\mathbf{\Omega},t)$ for different time intervals $\Delta t$.   (a)  Full range of scattering angles; (b) Diffraction
region. Calculations have been performed for the same parameters as Fig. 1.
 } \label{f2}
\end{figure}

Fig. 2 demonstrates the essential angular dependence, especially near the forward and backward directions. Such dependence takes place for all considered time intervals. For the
short time after the excitation pulse is switched off the superradiance is observed for radiation emitted in any directions (solid line).  For the very beginning of the
fluorescence the sideways scattering is characterized by a faster decay than forward one.

The maximal decay rate corresponds to an angle which depends on the size of the system (see below). Beginning with some time, the decay rate changes the sign for definite
angular intervals. This means that for corresponding time and angular intervals the intensity of fluorescence increases. Here we see manifestation of oscillation in the
afterglow of the atomic ensemble connected with quantum beating and caused by interference of light scattering through different collective states (see \cite{5,6,10}).

With time the spatial distribution of excited atoms in the cloud changes and the transverse distribution of emittance of the cloud changes as well. It causes modification of the
image of the diffraction.  In Fig. 2b we show the angular dependence of the decay rate for the region of the diffraction pattern on a large scale. One can see that the
diffraction picture transforms with time. Particularly, the separation between pairs of diffractions peaks changes. In our view, the transformation of the diffraction pattern is
responsible for the unusual angular dependence shown in Fig. 2. The intensity in a given direction changes not only because of decay of collective states but also because of
alteration in the direction of emission. The maximal decay rate is observed in directions of diffraction minima.

\begin{figure}[th]
\begin{center}
\includegraphics[width=18pc]{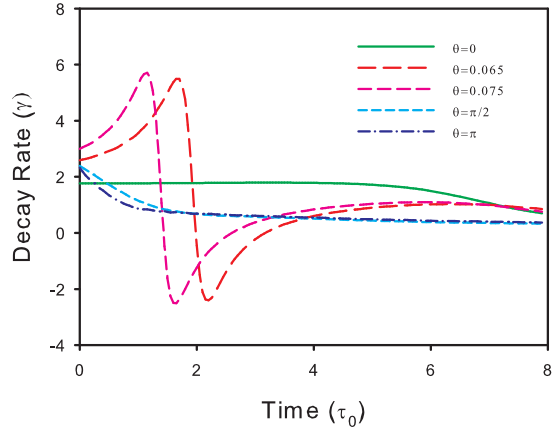}
\end{center}
\caption{Time dependence of current decay rate of the fluorescence in different directions.  Calculations have been performed for the same parameters as Fig. 1 and Fig. 2.
 } \label{f3}
\end{figure}

The difference in decay rates (\ref{10}) for different directions of fluorescence is shown also in Fig. 3. Just after the end of the exciting pulse ($t=0$)
$\Gamma(\mathbf{\Omega},0)>\gamma$ for any direction. As time passes $\Gamma(\mathbf{\Omega},t)$ changes. The afterglow into the  forward direction maintains a high value for
the  longest period of time. For $t$ up to $t=4\tau_0$ its value practically coincides with $\Gamma(0,0)$. It means that for this period of time the contribution of the
timed-Decay state into the forward emission is dominant. The decay rate of radiation into the backward half-sphere ($\theta>\pi/2$) decrease monotonically and for the considered
condition it loses its superradiant properties for $t\geq \tau_0$. Fluorescence into diffraction minima and higher order maxima demonstrate nonmonotonic, oscillatory behavior.
Two curves for $\theta=0.065$ and $\theta=0.075$ show that the decay rate can increase several times during the afterglow as well as decrease up to relatively large negative
values.

\begin{figure}[th]
\begin{center}
\includegraphics[width=18pc]{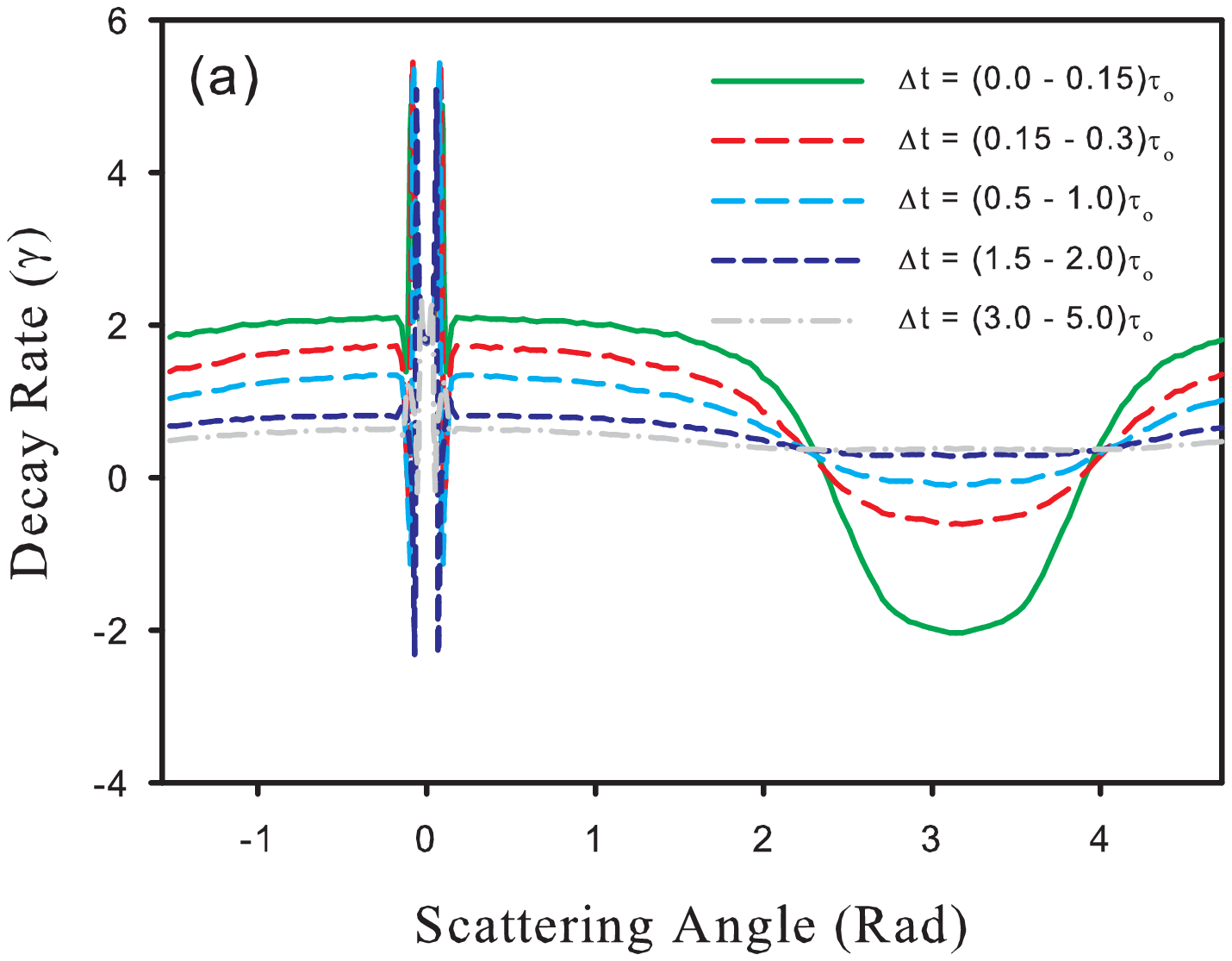}
\includegraphics[width=18pc]{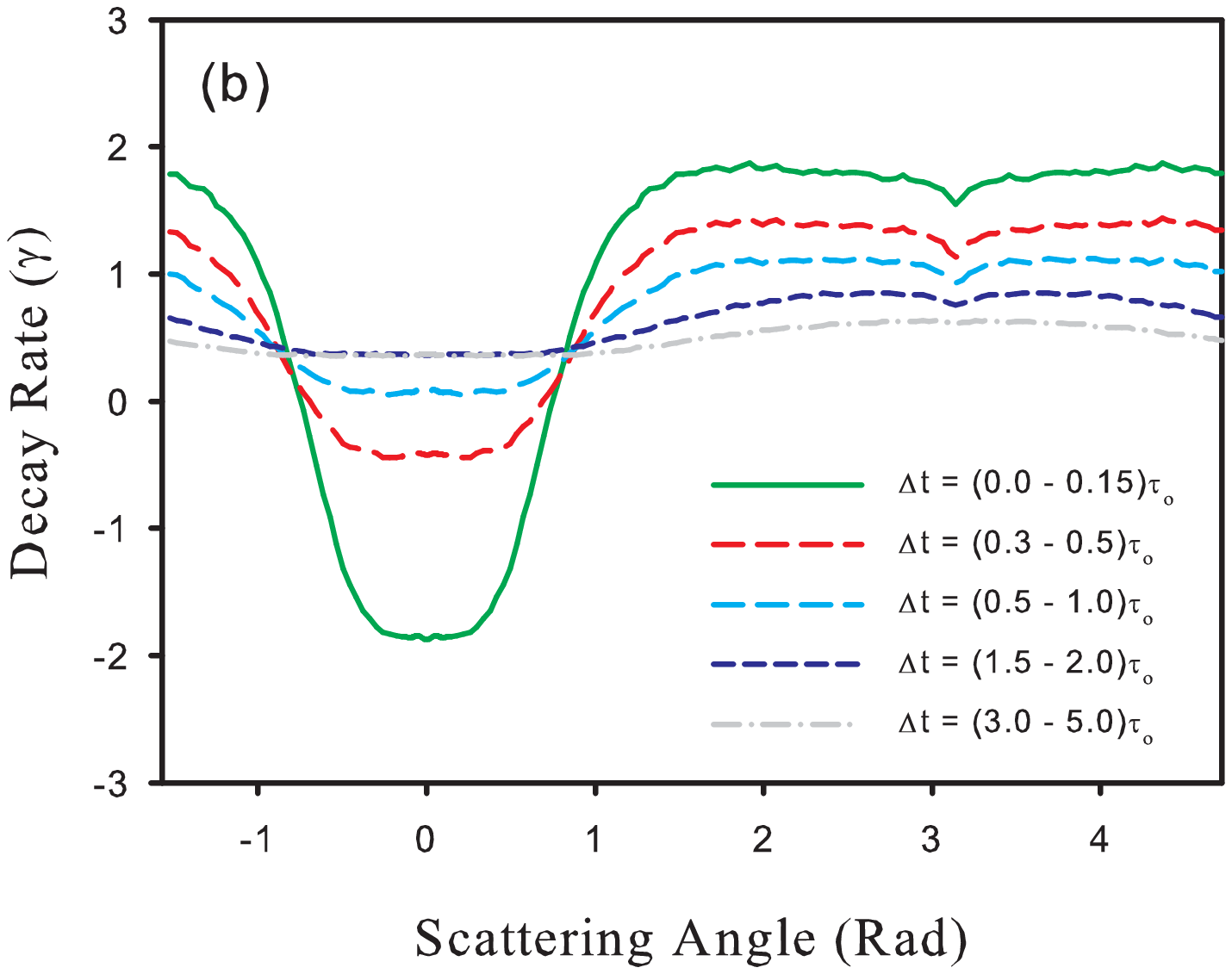}
\end{center}
\caption{Angular distribution of averaged decay rate $\Gamma_\alpha(\mathbf{\Omega},t)$ for different polarization channels.   (a) $H\| H$; (b) $H\bot H$. $\tau_L=\tau_0$. The
other parameters are as in Fig. 1.
 } \label{f4}
\end{figure}

The dynamics of the fluorescence in different polarization channels (see Fig. 4)  is even more complicated than that of the  total light intensity. It is connected with the
absence of single scattering in these channels into some specific direction. That is why the light intensity increases just after the pulse ends in the forward direction for the
case of $H\bot H$ and for the backward direction for $H\| H$.  In these directions the decay rate (\ref{10}) is negative and is relatively large in absolute value. With time the
contribution of high order scattering increases and we observe the usual decreasing of decay rate.

\subsection{Dependence on type of excitation}

\subsubsection{Dependence on the excitation duration}

In many cases, in theoretical papers devoted to single-photon superradiance the decay of the timed-Dicke state is discussed. However excitation of a real physical system into
such a state is a separate and difficult problem. The authors of \cite{5} found an original way to do so. In more traditional experiments with cold gases the excitation is
performed by means of pulsed radiation. The length of the pulse strongly influences the type of prepared atomic states and the consequent  fluorescence. In the case of a short
pulse many different collective states in a wide spectral region are excited and the type of decay may differ essentially from the decay of a timed-Dicke state. In the case of a
very long pulse we have a quasi steady state distribution of atomic excitation which differs from the distribution of the Dicke state. For the resonant excitation it is
connected with absorption and for a nonresonant one with dispersion \cite{14,15}. The spatial distribution of phases of the atomic oscillators is determined not by the wave
number of the exciting light but by the light wavelength in the medium.

The dependence of the nature of decay on the type of excitation for the forward scattering was discussed earlier (see for example \cite{5,11,21,22}). We will focus our attention
on its influence on the angular distribution of single-photon superradiance. The corresponding dependence for the most interesting angular region is shown in Fig. 5. We
demonstrate the angular dependence of decay rate both for very short and very long pulses.  For the exact forward direction we see the weakest changes when the length of the
pulse changes. It is connected with the fact that short lived collective states responsible for forward scattering have a larger width and they are effectively excited
independently of pulse length (spectral width of the pulse). At the same time, as it is seen from Fig. 6, the length of the pulse influences the duration of the process of
superradiance. As $\tau_L$ increases the duration grows shorter. This arises through the increasing role of sub-radiant collective states which are not practically shifted in
frequency. These states have a relatively small width and are weakly excited by a short pulse.
\begin{figure}[th]
\begin{center}
\includegraphics[width=18pc]{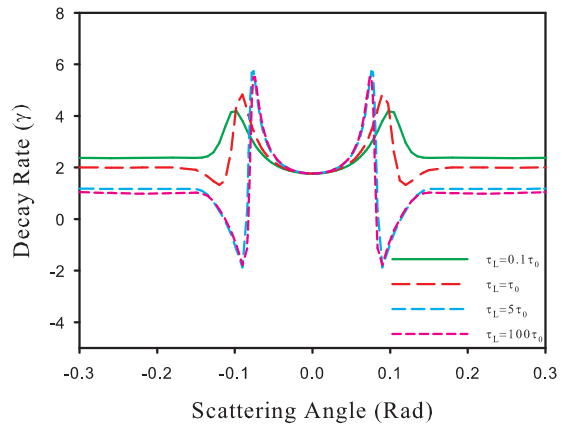}
\end{center}
\caption{Angular distribution of the decay rate for fluorescence excited by pulses of different length. Calculations have been performed for the same parameters as Fig. 1.
Observation time interval is $\Delta t=(0-0.1)\tau_0$
 } \label{f5}
\end{figure}
\begin{figure}[th]
\begin{center}
\includegraphics[width=18pc]{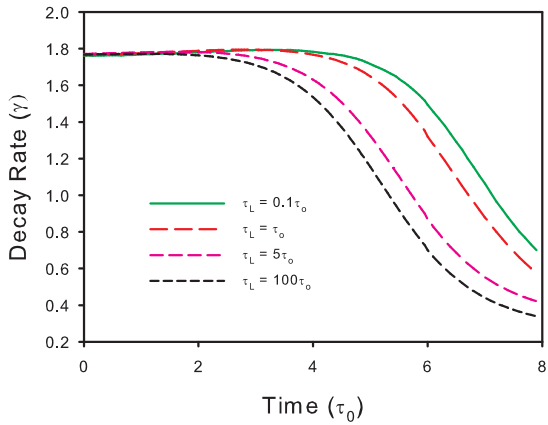}
\end{center}
\caption{Time dependence of the decay rate of fluorescence in the forward direction $\theta=0$. Calculations have been performed for the same parameters as Fig. 5.
 } \label{f6}
\end{figure}

For the directions where we observe a sharp angular dependence the length of the pulse influences at the very beginning of fluorescence. Because of the spatial inhomogeneity of
Gaussian clouds  the length of the resonant pulse causes transformation of not only longitudinal (along light propagation) but also the transverse distribution of excited atoms.
In its turn it causes transformation of the diffraction patterns. This is illustrated by Fig. 5.
\begin{figure}[th]
\begin{center}
\includegraphics[width=18pc]{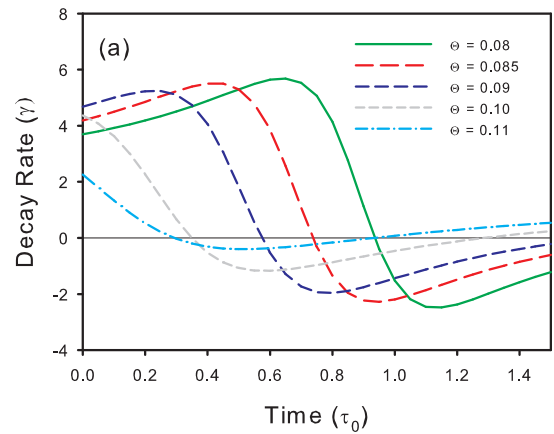}
\includegraphics[width=18pc]{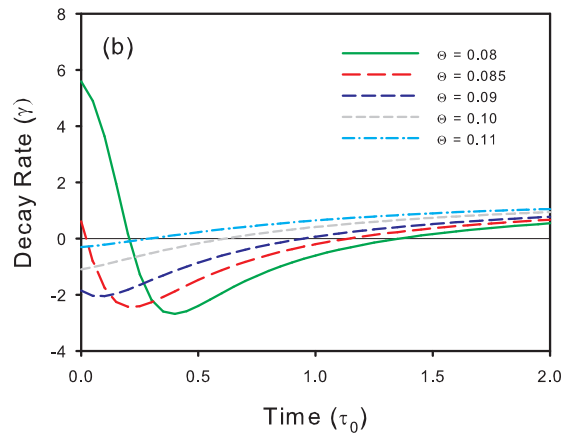}
\end{center}
\caption{Time dependence of the decay rate of fluorescence in different directions $\theta$. a) $\tau_L=0.1\tau_0$; b) $\tau_L=100\tau_0$.  The other parameters are as in Fig.1.
 } \label{f7}
\end{figure}

Besides changes in  width of the diffraction pattern we see a qualitative difference in angular dependence. For $\tau_L=0.1\tau_0$ there is only relatively small maxima whereas
already for $\tau_L=5\tau_0$ the maxima become more sharp, their amplitudes decreases essentially with angles and in some regions the decay rate changes sign. For further
increasing of $\tau_L$ we observe only small quantitative changes. A saturation type effect takes place. Increasing $\tau_L$ from $\tau_L=5\tau_0$ up to $\tau_L=100\tau_0$
practically changes neither intensity of fluorescence nor its rate. It means that for the considered cloud for $t=5\tau_0$ quasi static regime is realized.

The influence of excitation duration on fluorescence is also demonstrated in Fig. 7. Here we show the time dependence of the decay rate of fluorescence in different directions
for two pulse lengths $\tau_L=0.1\tau_0$ and $\tau_L=100\tau_0$. The qualitative difference between short (a) and long (b) excitation is that for a long pulse there are
directions for which intensity begins to increase immediately after the end of the pulse. Fig. 7 shows also that the excitation duration changes the nature of quantum beating.

\subsubsection{Single-photon superradiance for incoherent excitation}

The features of decay of a timed-Dicke state is essentially connected with phase matching of different atomic oscillators in the ensemble. Superradiance beyond the diffraction
zone is caused by incoherent scattering. In this connection the question arises whether it is necessary to use coherent excitation to observe sideward superradiance. Or it is
possible to observe this superradiance for incoherent excitation for complete absence of phase correlation. We performed calculation of atomic fluorescence assuming that
different atoms in the Gaussian cloud are excited independently. In such a case the phases of different atoms are random and average atomic polarization is absent.

Analyzing the time dependence of the fluorescence we calculated the current decay rate for an exciting pulse of different lengths. The results are shown in Fig. 8.
\begin{figure}[th]
\begin{center}
\includegraphics[width=18pc]{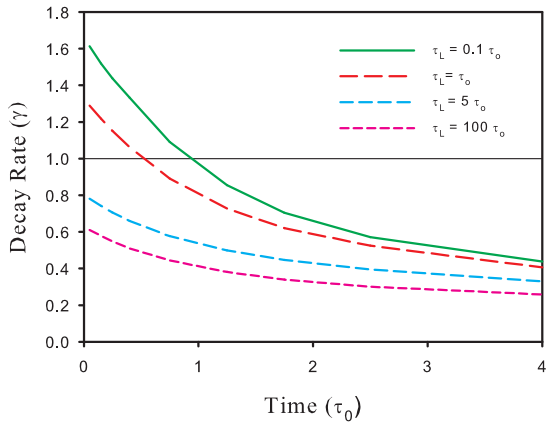}
\end{center}
\caption{Time dependence of the decay rate of fluorescence in the case of incoherent excitation by radiation of different pulse lengths. The other parameters are as in Fig. 1.
 } \label{f7}
\end{figure}

The calculation was made for the case when the carrier frequency of the pulse coincides with the resonant frequency of the free atoms. It is seen that for short pulses we
observe a decay with a  rate that exceeds the decay rate of the free atom i.e. we observe superradiance. For long resonant pulses the superradiance is absent. However, in the
paper \cite{7} it was shown that long coherent nonresonant  excitation causes superradiance of fluorescence in sideways directions. For this reason we studied how the decay rate
depends on carrier frequency of the radiation in the case of incoherent excitation.

In the following picture we show the corresponding dependence for two lengths of the pulse. The calculation is made for a spherically symmetric Gaussian cloud with radius $R=25$
and peak density $n=0.005$. Because of spherical symmetry (on average) of the cloud and excitation the secondary radiation of the atomic ensemble is spherically symmetric.
\begin{figure}[th]
\begin{center}
\includegraphics[width=18pc]{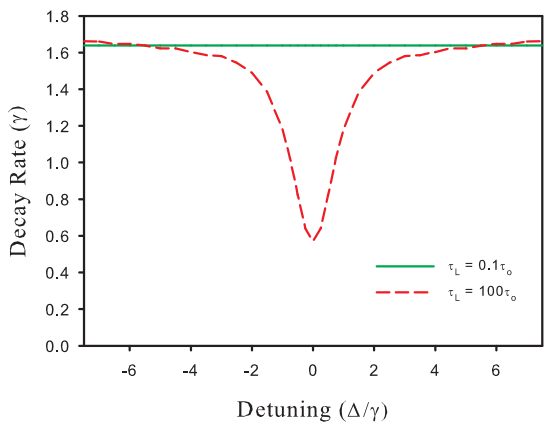}
\end{center}
\caption{Spectral dependence of the decay rate of fluorescence excited by incoherent radiation of different pulse durations.  The decay rate is calculated for the time interval
$\Delta t=(0-0.1)\tau_0$ after the end of the exciting pulse. The other parameters are as in Fig. 1.
 } \label{f8}
\end{figure}
For the short pulse $\tau_L=0.1\tau_0$, and because of its large spectral width the decay rate does not practically depend on carrier frequency. On the contrary for long pulse
$\tau_L=100\tau_0$ we see an essential dependence and, like in the case of coherent excitation \cite{7}, increasing of detuning causes increasing of decay rate up to some
constant magnitude which depends on the size of the cloud, density and which exceeds $\gamma$. The typical region of essential alteration of the rate is about the natural width
of the transition of the free atom. For this spectral region the optical depth of the cloud is big and for long pulses we have a quasi steady state regime of atomic excitation.
The atomic excitation under such conditions is determined not only by the external radiation but also by trapped light \cite{15}. By the time of the end of the exciting pulse,
the fluorescence is determined by photons scattered a different number of times inside the medium. For optically dense media scattering of high order plays an important role.
After the end of the pulse we still see contributions of different order scattering but only the single scattering is responsible for superradiance. The contributions of higher
order decay are much slower. That is why we do not see superradiance for long resonant pulse excitation. For nonresonant light the optical thickness is small and single
scattering in the sideways direction gives the main contribution and superradiance can be seen.

\subsection{Single-photon superradiance for ensembles of different sizes}

In this section we consider the dependence of angular distribution of superradiance on the sizes of the atomic ensemble. We will analyze both dependence on longitudinal and
transverse sizes.
\begin{figure}[th]
\begin{center}
\includegraphics[width=19.5pc]{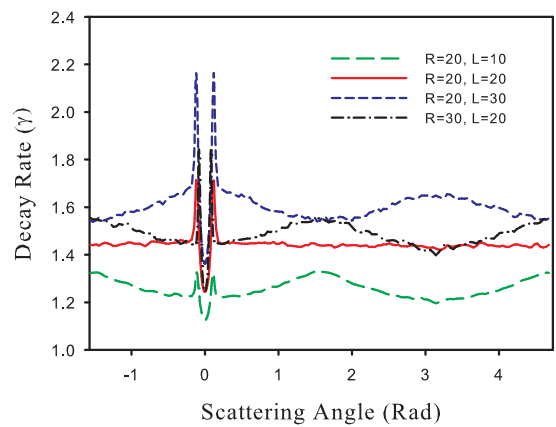}
\end{center}
\caption{Spectral dependence of the decay rate of fluorescence excited by incoherent radiation of different pulse duration.  Decay rate is calculated for the time interval
$\Delta t=(0-0.1)\tau_0$ after the end of the exciting pulse. $n_0=0.002.$ Pulse length is $\tau_L=0.1\tau_0$.
 } \label{f10}
\end{figure}

Let us consider at first how the angular distribution of the decay rate changes with length $L$ of the Gaussian clouds for fixed transverse radius $R$.  Results of the
corresponding calculations for several $L$ and $R=20$ are shown in the Fig. 10. The density in the center of the atomic ensemble is equal to $n=0.002$. The decay rate increases
with $L$ for all directions. But the specific dependence is different for different angles $\theta$.

For the forward direction we have
\begin{equation}
\label{21} \Gamma(\theta=0,t\rightarrow 0)=\gamma\left(1+\frac{b_{0z}}{8}\right),
\end{equation}
where $b_{0z}=\sqrt{2\pi}\sigma_0n_0L$ is the maximal resonant thickness of the Gaussian cloud along the $z$ direction. This expression coincides with results obtained earlier
in \cite{22} if we take into account that the authors of \cite{22} considered a scalar model of the radiation which underestimates the optical depth of the cloud by the factor
1.5.

The scattering in the near backward direction requires special attention. We see that for oblong clouds the decay rate into this direction ($\theta=\pi$) even exceeds that for a
timed-Dicke one ($\theta=0$). Note also that the decay rate for fluorescence at $\theta=\pi/2$ also increases in spite of the fact that the transverse size is fixed.

Such angular dependence as well as many important regularities of single photon superradiance can be understood in the framework of a random walk approach without analysis of
collective states of a polyatomic ensemble usually used in such a case. The random walk approach is very effective for description of incoherent multiple scattering in optically
thick but dilute media (see for example \cite{23}-\cite{26}). The light transport in a dilute medium generally performs a diffusion-type process, which in a semiclassical
picture can be visualized as a forwardly propagating wave randomly scattered by medium inhomogeneities. In an optically dense sample this process generates a zigzag-type path
consisting of either macroscopically or mesoscopically scaled segments of forwardly propagating waves. The forwardly propagating incoming, secondary and multiply scattered waves
can be expressed via a retarded-type Green's propagation function. This propagation function is completely described by the macroscopic susceptibility tensor. The incoherent
scattering events, which randomly happen in the medium, can be probabilistically simulated and properly described with the scattering theory formalism.

In practically important cases when sideward superradiance is observed, i.e. for short resonant or long nonresonant pulses the time dependence of the incoherent fluorescence,
just after the end of the exciting pulse, can be described by taking into account only single incoherent scattering.  The single scattering approximation is valid for a not very
big average optical depth $\overline{b_0}$ of the cloud  -- $\overline{b_0}\tau_0/\tau_L\lesssim 1$ (for short pulses) or $\overline{b_0}\tau_0\Delta_L\lesssim 1$ (for long
nonresonant pulse).  In such cases the intensity $I^s_\alpha(\mathbf{\Omega},t )$ can be calculated as follows
\begin{eqnarray}
\label{14}
&&I^s_\alpha(\mathbf{\Omega},t )=\int \frac{cn(\mathbf{r})}{4\pi\hbar^2} d^3 r\left\vert\int\limits_{-\infty }^{\infty }E_0(\omega)\dfrac{\exp(-i\omega t)d\omega }{2\pi } \right. \notag \\
&& k^{2} \chi(\mathbf{r},\mathbf{r}_a,\omega)\sum\limits_{e}\left. \frac{\left(\mathbf{u}^{\prime \ast }\mathbf{d}_{g;e}\right)
\left(\mathbf{ud}_{e;g}\right)}{\omega-\omega_a+i\gamma/2}
\chi(\mathbf{r}_a,\mathbf{r}_0,\omega)  \right\vert ^{2}. \notag\\
\end{eqnarray}
Here the function $\chi(\mathbf{r}_a,\mathbf{r}_0,\omega)$ describes propagation of light from the source to the point $\mathbf{r}_a$ where a single incoherent scattering event
takes place. The function $\chi(\mathbf{r},\mathbf{r}_a,\omega)$ describes propagation of a secondary photon toward the photodetector. In the isotropic medium these functions
are determined as follows
\begin{eqnarray}
\label{15} \chi(\mathbf{r}_2,\mathbf{r}_1,\omega)=\exp\left(-\frac{i b_0(\mathbf{r}_2,\mathbf{r}_1)}{2}\frac{\gamma/2}{\omega-\omega_a+i\gamma/2}\right) ,
\end{eqnarray}
where the resonant optical thickness of the inhomogeneous cloud between points $\mathbf{r}_1$ and $\mathbf{r}_2$ for considered case of $J=0\leftrightarrow J=1$ transition is
\begin{eqnarray}
\label{16}  &&b_0(\mathbf{r}_2,\mathbf{r}_1)=6\pi\lambdabar^2\int_{\mathbf{r}_1}^{\mathbf{r}_2}n(\mathbf{r})d\mathbf{s}.
\end{eqnarray}

Expression (\ref{14}) can be used for calculation of the decay rate for all directions except in the zones of backward and forward scattering. For forward scattering the main
contribution comes from the coherent component of the scattering light and for backward direction one of the polarization components is absent for single scattering and
scattering of higher order should be taken into account. Equation (\ref{14}) is also not valid for the cloud with a large aspect ratio. In such a case diffraction effects play
essential role  \cite{27,28} and the propagation function $\chi$ cannot be described by Eq. (\ref{15}).

The integral over frequency $\omega$ in (\ref{14}) can be calculated on the basis of the theory of residues. Restricting by the case of a typical experimental situation without
polarization analysis and taking into account that for rectangular pulse for $t>\tau_L$  only the pole $\omega=\omega_a-i\gamma/2$ is important, we have
\begin{eqnarray}
\label{19}
&&\Gamma(\mathbf{\Omega},t\rightarrow 0)=\gamma\left(1+\frac{\overline{b_0(\mathbf{r})}}{2}\right)=\\
&&\gamma\left(1+\frac{b_{0z}}{8}\left(1+ \frac{\sqrt{2}R}{\sqrt{R^2+L^2+(R^2-L^2)\cos(2\theta)}}\right)\right).\notag
\end{eqnarray}
Here  $b_0(\mathbf{r})$ is the total optical length of the resonant light ray coming from the source and incoherently scattered in the point $\mathbf{r}$ toward the detector
located along the direction $\theta$; $\overline{b_0(\mathbf{r})}$ is the value averaged over all atoms in the cloud.

In Fig. 11 we demonstrate the adaptability of Eq. (\ref{19}) for description of the angular distribution of the decay rate of single-photon superradiance. In this figure we
compare results of a quantum microscopic approach and approximate calculation of $\Gamma(\mathbf{\Omega},t\rightarrow 0)$ on the basis of Eq. (\ref{19}). It is clear that for
very small time interval $\Delta t=(0-0.01)\tau_0$ we have a good qualitative agreement. For bigger intervals $\Delta t=(0-0.1)\tau_0$ some quantitative discrepancy caused by
scattering of higher order appears, but qualitatively the angular dependence of $\Gamma(\mathbf{\Omega},t\rightarrow 0)$ is reproduced by the  Eq. (\ref{19}) quite well.
\begin{figure}[th]
\begin{center}
\includegraphics[width=19.5pc]{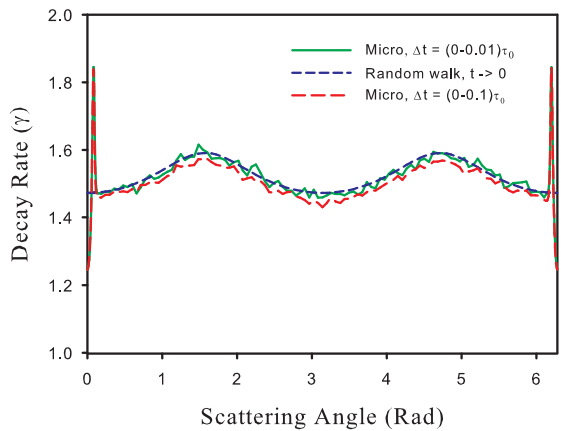}
\end{center}
\caption{Angular dependence of the decay rate of fluorescence. Comparison with analytical expression (\ref{19});   $\tau=0.01\tau_0$, $L=20$, $R=30$. The curves obtained in the
microscopic approach were calculated as a result of averaging over approximately 20000 different random spatial configurations of the atomic ensemble and were not smoothed. The
fluctuations in the curves demonstrates the accuracy of the calculations.
 } \label{f9}
\end{figure}

Note that in much the same way as we get Eq. (\ref{19}),  we can obtain expression (\ref{21}) for forward scattering if we consider only the coherent component of transmitted
light.

Good agreement between a microscopic approach and single scattering random walk approximation allows us to give a simple explanation of sideward superradiance in the considered
cases. The calculation based on Eq. (\ref{14}) shows that independently of the carrier frequency of the pulse the properties of secondary radiation after single scattering is
determined by optical depth of the cloud for quasi resonant radiation. Physically it is connected with the known fact that without external action any vibrating system
oscillates at its eigenfrequencies. For dilute media these frequencies are close to the free atom resonant frequency. Propagation of such quasi resonant radiation is accompanied by
spectral transformation. For not very big optical depth this propagation leads to broadening of the spectrum and consequently to acceleration of fluorescence. For large optical
depth distortion of the spectrum can be more essential. In such a case scattering of higher order should be taken into consideration.

\section{Conclusions}
In this paper we analyzed the time-dependent fluorescence of dilute Gaussian clouds of cold atoms excited by a weak quasi resonant light pulse. The calculation was performed on the basis of the quantum microscopic approach. Solving the nonstationary Schr\"{o}dinger equation for the joint system consisting of atoms and a weak electromagnetic field we
calculated the angular distribution and polarization properties of the fluorescence.

We focused our attention on the initial stage of fluorescence where superradiance was expected. Calculating transformation of the angular distribution of the afterglow of the ensemble with time we observed that for total emission without polarization analysis superradiance took place in any direction if the length of the pulse less or comparable with
natural lifetime of atomic excited states. Besides that there is substantial dependence of superradiance on the direction of fluorescence, especially in the region of the diffraction pattern and in the angular area of the coherent backscattering cone. Maximal decay rate is observed not for the forward direction but at some angle which is
determined by the transverse size of the cloud. This maximal value is several times more than the decay rate of the timed-Dicke state. Time-dependent fluorescence in separate polarization channels is more complicated. For a short exciting pulse there are directions where the corresponding polarization component does not decrease, but increases
initially. For example, for the helicity preserving channel it takes place for a direction close to the backward direction, for non preserving channels we see increasing of intensity into the forward direction.

For long coherent pulses the nature of fluorescence decay essentially depends on the frequency as was predicted in \cite{7}. For resonant radiation the superradiance is observed only in the forward direction. Moreover, there are directions near the main diffraction maximum where the total intensity summed over two orthogonal polarizations increases just
after the end of exciting pulse. As the carrier frequency shifts from exact atomic resonance the decay rate in sideward directions increases and for some detunings superradiance takes place (see also \cite{7}).

We repeated  analysis of  single-photon superradiance for the incoherent excitation and found that the superradiance can be observed in this case, i.e. when atomic polarization in the ensemble is absent. It can be excited either by a short pulse or by a long nonresonant one.

We studied the dependence of the angular distribution of superradiance on the size and shape of the atomic ensemble. Besides a sharp features connected with the diffraction pattern, and the coherent backscattering cone, we observed noticeable transformation of this dependence caused by changes in aspect ratio of the cloud. The decay rate is
determined by average optical depth of the cloud for singly scattered photon.

Besides a quantum microscopic approach we analyzed single-photon superradiance on the basis of random walk theory.  We showed that for not very big optical depth of the cloud, and  in the case of very short resonant and long nonresonant pulses, the time dependence of the incoherent fluorescence just after the end of the pulse can be described by
taking into account only single incoherent scattering. In such a case we derive simple analytical expressions for the decay rate of single-photon superradiance in an arbitrary direction.

This expression was obtained under the assumption of motionless atoms and for very simple atomic level structure. However it is quite clear that a random walk approach can be applied for the more general case, for example for moving atoms with hyperfine level structure. Scattering of any order also can be taken into account. In our opinion this
approach would be useful in the presence of a control field. Such a field changes the spectral properties of the medium essentially which strongly influences the incoherent scattering under conditions of electromagnetically induced transparency \cite{29,30} and may modify single-photon superradiance in cold and dilute atomic gases.

\begin{acknowledgments}
We appreciate financial support by the Russian Foundation for Basic Research (Grant No. RFBR-15-02-01013). A.S.K. also thanks RFBR-16-32-00587 and the Council for Grants of the
President of the Russian Federation.  We also acknowledge financial support by the National Science Foundation (Grant No. NSF-PHY-1606743).
\end{acknowledgments}

\bigskip

\end{document}